\documentclass{mpe_report}

\usepackage{psfig,graphicx,epsfig}
\usepackage{color}
\usepackage{amsmath,amssymb,epic,eepic,array}

\begin{document}

\pagenumbering{arabic}
\setcounter{page}{36}

\renewcommand{\FirstPageOfPaper }{ 36}\renewcommand{\LastPageOfPaper }{ 39}%%
%% This is MPE-Report_example.tex
%% LaTeX2e example style file for the contributed talks and posters presented
%% during the 363rd Heraeus Seminar on Neutron Stars and Pulsars, held in 
%% Bad Honnef, May 14.-19. 2006.
%% 
%% This file needs the LaTeX2e class file he_symp.cls  
%%
% -----------------------------------------------------------------------------
%\documentclass{mpe_report}
%\usepackage{psfig}
% -----------------------------------------------------------------------------
%\begin{document}

\title{Gamma-ray Pulsar Simulations for the Gamma-ray Large Area Space Telescope (GLAST)}
\author{M. Razzano\inst{1,2}} 
\date{July, 26 2006}
\institute{Istituto Nazionale di Fisica Nucleare, Pisa sect., Largo B.Pontecorvo 3, 56100 Pisa, Italy. email:massimiliano.razzano@pi.infn.it
\and \emph{on behalf of the GLAST LAT Collaboration}}
\maketitle

\begin{abstract}
We present here the current status of pulsar simulations for the Large Area Telescope, 
the main instrument on the GLAST mission. We present \emph{PulsarSpectrum}, a pulsar simulator
that can reproduce with high detail gamma-ray emission from pulsars.
\emph{PulsarSpectrum} takes into account advanced timing effects,
e.g. period changes with time, barycentering effects and glitches.
Other ancillary tools have been built to provide the simulator with
a realistic population of pulsars and their ephemerides. All these
tools are currently used in the GLAST collaboration for testing the
LAT Science Analysis Tools and for studying LAT capabilities for
pulsar science. They have been used also for generating a
simulated pulsar population for the  \emph{Data Challenge 2} (DC2), one of
the most important milestones in the development of the GLAST
software. During the DC2, scientists analyzed a set of 55 days of
simulated data in order to validate LAT MonteCarlo, study instrument
response functions, exercise analysis tools and study LAT
capabilities. This contribution also contains results of some
analysis performed during DC2 on EGRET pulsars.
\end{abstract}

\section{Introduction}
Pulsars are among the most exciting gamma-ray sources in the
Universe and can serve as unique sites for the study of emission
processes in extreme physical environments. The Gamma ray Large Area
Space Telescope (GLAST) will increase dramatically our knowledge of
gamma ray pulsars physics. In particular the Large Area Telescope
(LAT), the main GLAST instrument, will provide more detailed
observations of the known gamma-ray pulsars and potentially will
discover many new pulsars that emit gamma rays. To better understand
the capabilities of GLAST for pulsar science we developed {\em
PulsarSpectrum}, a program that simulates gamma ray emission from
pulsars and a set of ancillary tools that can create simulated pulsar data to be used with \emph{PulsarSpectrum}. This simulator can be easily interfaced with the MonteCarlo software that simulates the response of the LAT.\\
Today seven high-confidence and three low-confidence gamma-ray pulsars are known and this numbers will increase with the observations of GLAST, scheduled for launch in the fall of 2007.\\
\begin{figure}
\centerline{\psfig{file=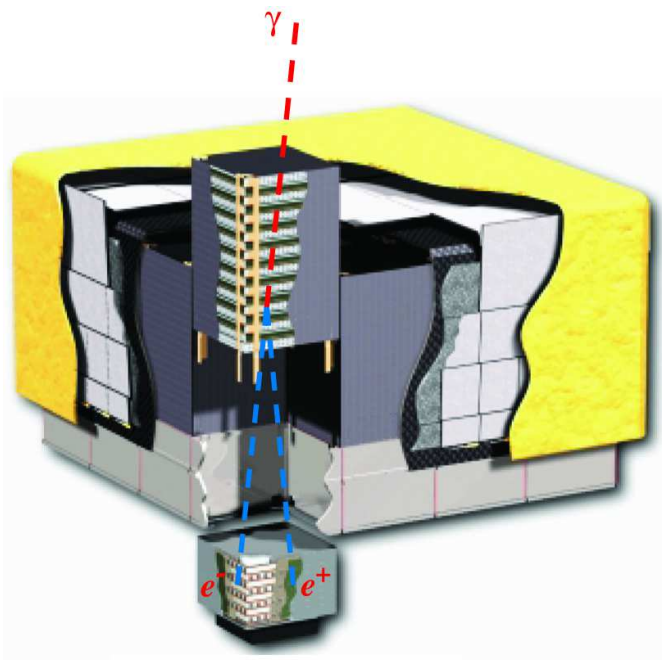,width=8.8cm,clip=} }
\caption{The
GLAST Large Area Telescope. An incident gamma ray converts
into a $e^{-}e^{+} $ pair
\label{lat}}
\end{figure}
The GLAST Large Area Telescope (LAT) (Fig. \ref{lat}), is a pair
conversion telescope based on the most advanced high energy
detectors. It consists in a precision silicon strip tracker, an
hodoscopic calorimeter and a segmented anticoincidence shield for
particle background rejection. The LAT high sensitivity (2$\times$10$^{-9}$
ph/cm$^{2}$/s) and effective area ($>$8000 cm$^{2}$) will permit the
discovery of a lot of new pulsars: the estimates range between tens
to hundreds depending upon the theoretical scenario adopted. Moreover the low dead
time of the detector (20 $\mu$s) will allow the detailed
reconstruction of pulsar lightcurves. One of the most exciting
possibilities of the LAT will be the coverage of the energy window
from 30 GeV up to 300 GeV, a still unexplored range. At these
energies the theoretical models make different previsions on the
high energy spectral cutoff and the spectral coverage of LAT will be
of primary importance for constraining and discriminating among the
models. In order to study the LAT response to specific gamma ray
sources, various simulation packages have been developed. 

\section{The {\em PulsarSpectrum} simulator}
\subsection{General overview}
The basic idea of {\em PulsarSpectrum} is to construct a
bidimensional histogram representing the differential flux vs.
energy and pulsar phase. This histogram contains all the basic
informations about lightcurve and spectrum. How it is built depends
upon the model we want to use: for example a phenomenological model, based only
on observations, or a physical one, based on a specific theoretical
model. Currently two models have been included.
The first model implemented was phenomenological, since it is more flexible and 
completely independent from the theoretical emission scenario. A second model recently implemented allow the user to simulate pulsars with an arbitrary photon distribution in phase and in energy. This model can in fact use a 2D histogram that can be provided by other external programs.\\
The input parameters of the simulator
can be divided in two categories and are listed in the Table \ref{PSParam}:

\begin{table}
      \caption{The \emph{PulsarSpectrum} parameters.}
         \label{PSParam}
      \[
         \begin{array}{p{0.5\linewidth}r}
            \hline
            \noalign{\smallskip}
            Type of Parameter    &  {\rm  Parameter} \\
            \noalign{\smallskip}
            \hline
            \noalign{\smallskip}
             Observational$^{a}$ :   & RA, Dec, flux,\\
                               & t_{0},f(t_{0}),f'(t_{0}),f''(t_{0}), \\
             Model-dependent:  & Model, E_{min}, E_{max},\\
                               & $Random seed,$ \\ 
                               & $5 free parameters,$\\
            \noalign{\smallskip}
             \hline
         \end{array}
      \]
\begin{list}{}{}
\item[$^{\rm a}$] t$_{0}$ indicates the \emph{epoch} of the ephemerides 
\end{list}
   \end{table} 

\begin{figure}
\centerline{\psfig{file=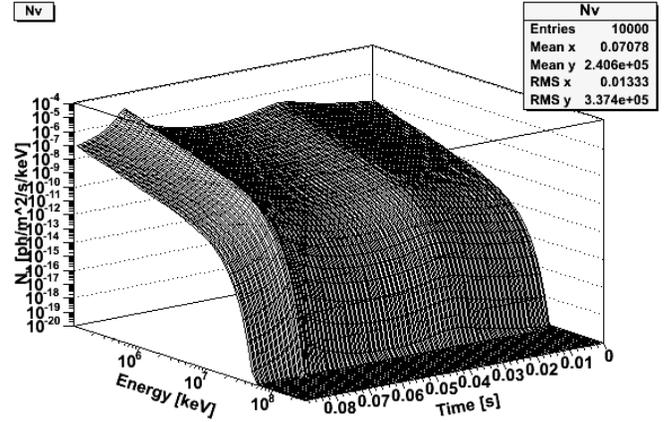,width=8.8cm,clip=} }
\caption{An example of a 2D histogram generated by \emph{PulsarSpectrum} for simulating a pulsar with a spectrum similar to Vela pulsar
\label{velanv}}
\end{figure}

\begin{itemize}
\item {\em Observational parameters}, which characterize the general parameters of the pulsar;
\item {\em Model-dependent parameters}, that define which model will be used for simulation and the set of parameters used by this model. There are 5 free parameters that leave to the user the freedom to adjust the model. 
\end{itemize}
All parameters are placed in two specific data files used by both
{\em PulsarSpectrum} and the LAT simulation tools. {\em
PulsarSpectrum} creates the lightcurve and the spectrum from these
parameters and combines them to obtain a two-dimensional matrix that
represents the flux in ph/m$^2$/s/keV. An example of such an histogram for a simulated pulsar similar to Vela is in Fig. \ref{velanv}.
The photons are then
extracted such that the interval between two subsequent photons is
determined by the flux integrated over the energy range of interest.
\subsection{The phenomenological model}
\begin{figure}
\centerline{\psfig{file=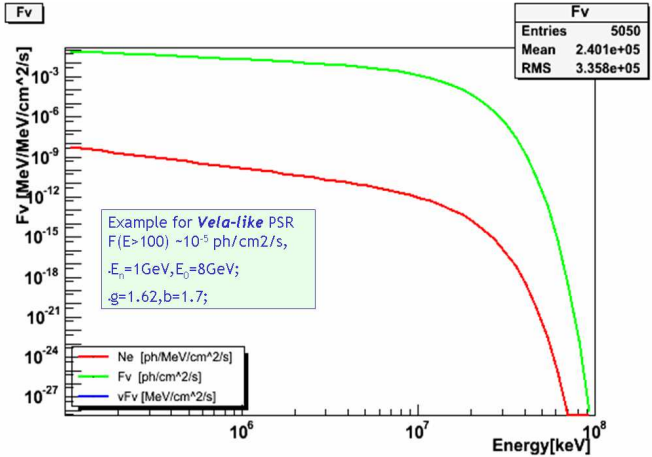,width=8.8cm,clip=} }
\caption{The input spectrum for a Vela-like pulsar using phenomenological model. The spectral parameters have been obtained from [\cite{NelDj95}]. The crosses indicate the distribution of the extracted photons.
\label{velaspec}}
\end{figure}

The phenomenological model allows the user to
generate pulsar lightcurves in a general way using a single or
double Lorenzian peak profile whose shape is determined from random
generated numbers. The lightcurve can be generated alternatively
from a user-provided profile. This is useful for simulating the
already observed gamma ray pulsars. The spectral shape is assumed to
be a power law with exponential cutoff (according to
[\cite{NelDj95}]), as in the observed gamma-ray pulsars and can
then be modeled as:
\begin{equation}
\frac{dN}{dE} =  K(\frac{E}{E_{n}})^aexp(\frac{E}{E_0})^{-b}
\end{equation}
The normalisation constant  K is determined by the photon flux in the range 
100 MeV- 30 GeV, in order to have flux compatible with the fluxes in the 3$^{rd}$ EGRET Catalog.
The other spectral parameters can be varied; the values for
the EGRET pulsars are obtained from real data by fitting procedures
(See e.g. [\cite{NelDj95}]).
As an example, in Fig. \ref{velanv} is presented an histogram for a \emph{Vela-like} pulsars, i.e. a simulated pulsar with characteristics similar to the Vela pulsar. The lightcurve has been derived from EGRET observations while the spectral parameters from a multiwavelength fit as reported in \cite{NelDj95}, are displayed in Fig. \ref{velaspec}.
\subsection{Timing issues}
Once the differential flux histogram is created the time interval
between two subsequent photons is computed according to the flux. The strategy adopted is to compute the mean photon rate and then to calculate the interval to the next photon according to Poisson statistics.
Then a secondary correction $\Delta$t is summed to this interval in order that the photon distribution follow the input lightcurve. 
The interval between two photons is computed assuming that the
pulsar period does not change with time and the photons arrival
times are computed into a reference system fixed relative to stars.
This is not the "real world". Pulsar timing is affected by more
complicate effects, as  (1)- The motion of GLAST through
the Solar System and the relativistic effects due to gravitational
well of the Sun (see \ref{barydecorr}); (2)- Period changes with
time (see \ref{pchange}). For pulsars in binary systems an
additional modulation to the orbital period should be taken into
account. For a precise pulsar simulator intent to produce a
realistic list of photon arrival times we need to include all these
effects (to transform to the observational frame). All these
procedures are now implemented in the code and only the binary
demodulation is not yet implemented. The real arrival time of a
photon from a pulsar must be first barycentered and then phase
assigned.
\subsubsection{Barycentric effects}\label{barydecorr}
The first step to analyze pulsar data is the conversion from the
arrival times at the spacecraft, usually expressed in Terrestrial
Time TT or TAI, to the arrival times at the Solar System barycenter,
expressed in Barycentric Dynamical Time TDB. Taking into account
both the motion of spacecraft through space and the general
relativistic effects due to the gravitational field of the Sun (i.e.
Shapiro delay), the simulator computes the opposite of the
barycentric correction by considering the position of the Earth and
of the spacecraft in the Solar System, and the position of the Sun.
The accuracy for the computation of these corrections is
hard-coded in the program.
\subsubsection{Period change and ephemerides}\label{pchange}
The rotational energy of a radio pulsar decreases with time and
hence the period increases with time. For gamma-ray pulsar science
the radio ephemerides are fundamental for assigning the correct
phase to each photon. If we know the frequency {\em f$(t_{0}$)} and
its derivatives {\em $\dot{f} (t_{0})$} and {\em $\ddot{f} (t_{0})$}
at a certain time t$_{0}$, known as {\em epoch}, the phase is then:
\begin{equation}\label{phit}
\phi(t) = int[ f(t_{0})(t-t_{0}) + \frac{1}{2}\dot{f}
(t_{0})(t-t_{0})^{2} + \frac{1}{6}\ddot{f} (t_{0})(t-t_{0})^{3}].
\end{equation}
The interval between two photons must be also corrected for this
effect. In the parameters file the user can specify a set of
ephemerides with the relative epoch of validity expressed in
Modified Julian Date. The simulator then computes the opportune
arrival time such that, after applying the barycentric corrections
and then the Eq. \ref{phit}, the correct phase is obtained.
\begin{figure}
\centerline{\psfig{file=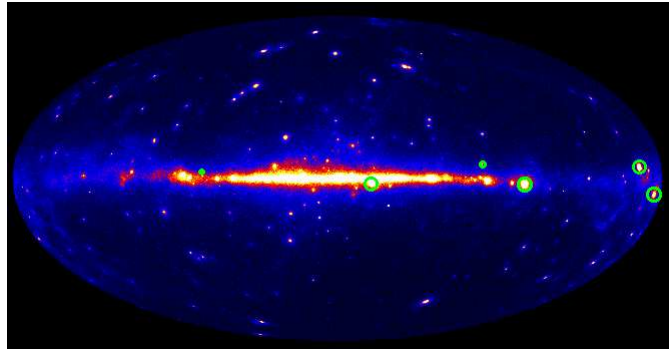,width=8.8cm,clip=} }
\caption{The count map of the simulatd data of the LAT DC2. The green circles indicate the EGRET pulsars, part of the whole pulsar population simulated using \emph{PulsarSpectrum}
\label{dc2sky}}
\end{figure}
\section{Gamma-ray pulsars in the LAT Data Challenge 2}
\begin{figure}
\centerline{\psfig{file=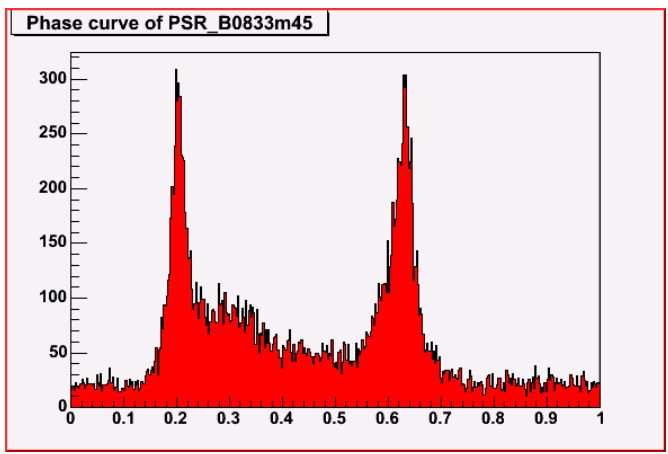,width=8.8cm,clip=} }
\caption{The phase distribution of the photons above 100 MeV from the Vela pulsar using DC2 simulated data. 
\label{veladc2}}
\end{figure}

\begin{figure}
\centerline{\psfig{file=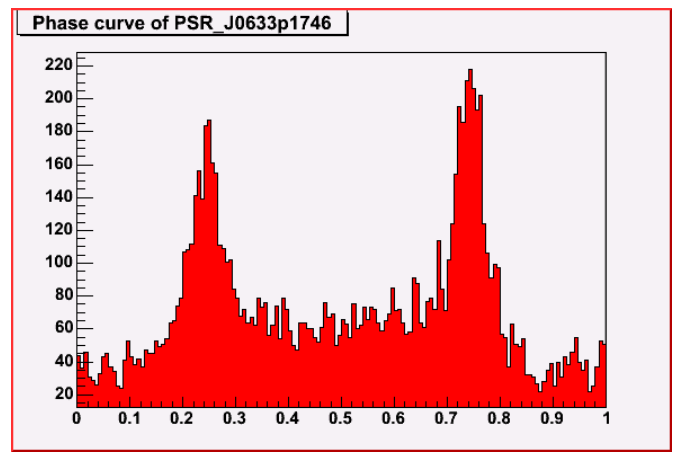,width=8.8cm,clip=} }
\caption{The phase distribution of the photons above 100 MeV from the Geminga pulsar using DC2 simulated data.\label{gemindc2}}
\end{figure}

\emph{PulsarSpectrum} have been used by the GLAST LAT Collaboration for simulating pulsars in the LAT \emph{Data Challenge 2}. The Data Challenge 2 (DC2) was an important milestone for the preparation of the 
analysis and simulation software for the GLAST mission. For the DC2 a simulated 55-days long LAT observation of the whole sky have been generated, and the simulated data have been provided to the scientists of the Collaboration. The main goals of the DC2 were to study and validate the LAT MonteCarlo, to exercize the Analysis Tools under development for GLAST and to better study the LAT Instrument Response Functions. Using DC2 simulated data, a group of LAT scientists have worked on these from the begin of March to the end of May (date of official DC2 Closeout meeting) and beyond. \\
One of the most important efforts in the DC2 was the composition of an high-detailed simulation of the gamma-ray sky (see Fig. \ref{dc2sky}). This similated sky was built by using a set of programs capable to simulate the different gamma-ray sources. Not only the current known gamma-ray sources discovered by EGRET have been included, but also a lot of new sources, in order to push down the sensitivity limit of the detected sources. Also possible new classes of gamma-ray sources have been included, such X-ray binaries and microquasars.\\
The pulsar component was entirely simulated by \emph{PulsarSpectrum} and was composed by some subclasses of pulsars. There are the six gamma-ray pulsars detected at EGRET energies, displayed in Fig. \ref{dc2sky}, and some other real radio-pulsars that are within error box of some 3EG sources. The remaining pulsars were not actually real pulsars, but simulated pulsars that have characteristics similar to the current known pulsars. These were divided in two subgroups, a set of isolated pulsars and a set of millisecond pulsars. A database containing the timing solution for some pulsars has been provided to DC2 users. However for some gamma-ray pulsars the timing solution have not been provided, in order to leave to the DC2 people the possibility to study tecniques for finding \emph{Geminga-like} pulsars with no radio counterparts.
In Fig. \ref{veladc2} and \ref{gemindc2} are displayed the recontructed phase curves of the simulated Vela and Geminga pulsar using DC2 simulated data.
These curves showed a profile very similar to the real EGRET profile, but this is not a suprise since the EGRET data have been used for shaping the input lightcurve for EGRET pulsars present in the DC2. With comparison of EGRET it is remarkable the better statistics achieved with the LAT in 55 days in comparison to the EGRET curves, as reported e.g. in [\cite{Kanbach02}]
\section{Conclusions}
Pulsar simulations are very important to study the GLAST LAT capabilities to probe these gamma-ray sources. We present here the latest results in pulsar simulations using the \emph{PulsarSpectrum} simulator, able to reproduce timing and spectral features of gamma-ray pulsars with high detail. 
{\em PulsarSpectrum} has been successfully used by the GLAST
collaboration for testing the functionality of the LAT Science
Analysis Tools, a set of analysis programs specifically designed to
analyse the LAT data after launch. An important event in the preparation of the GLAST mission was the Data Challenge 2, during what a 55-days LAT observation of the full sky have been generated. Pulsars included in DC2 sky have been simulated with this simulator, providing LAT scientists simulated data from a realistic gamma-ray pulsar population. The DC2 has been also a good chance to exercize the pulsar simulator and to show that \emph{PulsarSpectrum} can be used by the GLAST Collaboration for better exploring the LAT detecting capability on pulsars.

\vskip 0.4cm   
% Example list of References

%\end{document}

          \clearpage

\end{document}